\documentclass{elsart}

\usepackage{graphicx}

\usepackage{amssymb}

\journal{Physica A}

\begin{document}

\begin{frontmatter}



\title{The Sherrington-Kirkpatrick spin glass model in the presence of
a random field with a joint Gaussian probability density function
for the exchange interactions and random fields}


\author{Ioannis A. Hadjiagapiou\corauthref{cor1}}
\corauth[cor1]{Corresponding author.} \ead{ihatziag@phys.uoa.gr}

\address{Section of Solid State Physics, Department of Physics,
University of Athens, Panepistimioupolis, GR 15784 Zografos,
Athens, Greece}

\begin{abstract}

The magnetic systems with disorder form an important class of
systems, which are under intensive studies, since they reflect
real systems. Such a class of systems is the spin glass one, which
combines randomness and frustration. The Sherrington-Kirkpatrick
Ising spin glass with random couplings in the presence of a random
magnetic field is investigated in detail within the framework of
the replica method. The two random variables (exchange integral
interaction and random magnetic field) are drawn from a joint
Gaussian probability density function characterized by a
correlation coefficient $\rho$. The thermodynamic properties and
phase diagrams are studied with respect to the natural parameters
of both random components of the system contained in the
probability density. The de Almeida-Thouless line is explored as a
function of temperature, $\rho$ and other system parameters. The
entropy for zero temperature as well as for non zero temperatures
is partly negative or positive, acquiring positive branches as
$h_{0}$ increases.

\end{abstract}

\end{frontmatter}

\newpage
\section{Introduction}

The critical properties of magnetic systems with quenched disorder
has been a topic of growing interest in statistical physics over
the last years and still attracts considerable attention in spite
of their long history, since it has been established that the
introduction of randomness can cause important effects on their
thermodynamic behavior in comparison to the pure ones; the type of
the phase transition as well as the universality class may change
\cite{stinchcombe,dotsenkodotsenko,imrywortis}. In two-dimensions
an infinitesimal amount of disorder converts a second-order phase
transition (SOPT) into a first-order phase transition (FOPT),
whereas in three-dimensions it is converted to an FOPT only when
disorder exceeds a threshold. Randomness is encountered in the
form of vacancies, variable or diluted bonds, impurities
\cite{hadjietal,hadjiag}, random fields
\cite{hadjiagapioubin,fytas1,fytas2,arabs,physicstoday,imryma,aharony,nishimortiz}
and spin glasses
\cite{nishimortiz,binderyoung,binder,nishim1,mezardetal,fischer,uspekhi}.
In addition, the study of disordered systems is a necessity
because homogeneous systems are, in general, an idealization,
whereas real materials contain impurities, nonmagnetic atoms or
vacancies randomly distributed within the system consisting of
magnetic atoms, or variable bonds in magnitude and/or in sign;
such systems have attracted wide interest and have been studied
intensively theoretically, numerically and experimentally, since
it is a matter of great urgency to develop an understanding of the
role of these non ideal effects. However, to facilitate their
study, these are modeled by relying on pure-system models modified
accordingly, e.g. Ising, Potts, Baxter-Wu, etc.

An important manifestation of disorder is the presence of random
magnetic fields acting on each spin of the magnetic system under
consideration in an otherwise free of defects lattice, whose pure
version is modeled according to a current one, e.g. the Ising
model; the system now in the presence of such fields is called
random field Ising model (RFIM)
\cite{hadjiagapioubin,fytas1,fytas2,arabs,physicstoday,imryma,aharony,nishimortiz}.
Associated with this model are the notions of lower critical
dimension, tricritical points, scaling laws, crossover phenomena,
higher order critical points and random field probability
distribution function (PDF). RFIM had been the standard vehicle
for studying the effects of quenched randomness on phase diagrams
and critical properties of lattice spin systems and had been
studied for many years since the seminal work of Imry and Ma
\cite{imryma}. The RFIM Hamiltonian, in the case of constant
exchange integral, is

\begin{equation}
 H=-J\sum_{<i,j>}S_{i}S_{j}-\sum_{i}h_{i}S_{i}  \hspace{2mm},
    \hspace{20mm} S_{i}=\pm1 \label{rham}
\end{equation}

This Hamiltonian describes the competition between the long-range
order (expressed by the first summation) and the random ordering
fields. We also consider that $J > 0$ so that the ground state is
ferromagnetic in absence of random fields. The RFIM model, despite
its simple definition and apparent simplicity, together with the
richness of the physical properties emerging from its study, has
motivated a significant number of investigations; however, these
properties have proved to be a source of much controversy,
primarily due to the lack of a reliable theoretical foundation.
Still, substantial efforts to elucidate the basic problems of the
RFIM continue to attract considerable attention, because of its
direct relevance to a number of significant physical problems. The
presence of random fields requires two averaging procedures, the
usual thermal average, denoted by angular brackets
$\langle...\rangle$, and disorder average over the random fields
denoted by $\langle...\rangle_{r}$ for the respective PDF, which
is usually a version of the bimodal, trimodal or Gaussian
distributions. The most frequently used PDF for random fields is
either the bimodal or the single Gaussian; the former is,

\begin{equation}
 P(h_{i})=p\delta(h_{i}-h_{0}) + q \delta (h_{i}+h_{0})  \label{bimodal}
\end{equation}

where $p$ is the fraction of lattice sites having a magnetic field
$h_{0}$, while the rest sites have a field $(-h_{0})$ with site
probability $q$ such that $p+q=1$; the usual choice was $p = q =
\frac{1}{2}$, symmetric case. The latter PDF is,

\begin{equation}
P(h_{i}) = \frac{1}{(2 \pi \sigma ^{2})^{1/2}} \; exp\left[-
\frac{h^{2}_{i}}{2 \sigma ^{2}}\right]   \label{sexp}
\end{equation}

with zero mean and standard deviation $\sigma$.

One of the main issues was the experimental realization of random
fields. Fishman and Aharony \cite{fishaha} showed that the
randomly quenched exchange interactions Ising antiferromagnet in a
uniform field $H$ is equivalent to a ferromagnet in a random field
with the strength of the random field linearly proportional to the
induced magnetization. This identification gave new impetus to the
study of the RFIM, the investigation gained further interest and
was intensified resulting in a large number of publications
(theoretical, numerical, Monte Carlo simulations and experimental)
in the last thirty years. Although much effort had been invested
towards this direction, the only well-established conclusion drawn
was the existence of a phase transition for $d \geq 3$ (d space
dimension), that is, the critical lower dimension $d_{l}$ is 2
after a long controversial discussion \cite{imryma,imbrie}, while
many other issues are still unanswered; among them is the order of
the phase transition (first or second order), the universality
class and the dependence of these points on the form of the random
field PDF. Galam, via MFA, has shown that the Ising
antiferromagnets in a uniform field with either a general random
site exchange or site dilution have the same multicritical space
as the random-field Ising model with bimodal PDF \cite{galam3}.
The study of RFIM has also highlighted another feature of the
model, that of tricriticality and its dependence on the assumed
distribution function of the random fields. A very controversial
issue has arisen concerning the effect of the random-field
probability distribution function on the equilibrium phase diagram
of the RFIM. The choice of the random field distribution can lead
to a continuous ferromagnetic/paramagnetic (FM/PM) boundary as in
the single Gaussian PDF, whereas for the bimodal one this boundary
is divided into two parts, an SOPT branch for high temperatures
and an FOPT branch for low temperatures separated by a tricritical
point (TCP) at $kT^{t}_{c}/(zJ)=2/3$ and
$h^{t}_{c}/(zJ)=(kT^{t}_{c}/(zJ)) \times
\arg\tanh(1/\sqrt{3})\simeq 0.439$ \cite{aharony}, where $z$ is
the coordination number and $k$ the Boltzmann constant, such that
for $T<T^{t}_{c}$ and $h>h^{t}_{c}$ the transition to the FM phase
is of first order for $p = \frac{1}{2}$. However, this behavior is
not fully elucidated since in the case of the three-dimensional
RFIM, the high temperature series expansions by Gofman et al
\cite{gofman} yielded only continuous transitions for both
probability distributions, whereas according to Houghton et al
\cite{houghton} both distributions predicted the existence of a
TCP, with $h^{t}_{c} = 0.28 \pm 0.01$ and $T^{t}_{c} = 0.49\pm
0.03$ for the bimodal and $\sigma^{t}_{c} = 0.36 \pm 0.01$ and
$T^{t}_{c} = 0.36\pm 0.04$ for the single Gaussian. In the Monte
Carlo studies for $d = 3$, Machta et al \cite{machta}, using
single Gaussian distribution, could not reach a definite
conclusion concerning the nature of the transition, since for some
realizations of randomness the magnetization histogram was
two-peaked (implying an SOPT) whereas for other ones three-peaked,
implying an FOPT; Middleton and Fisher \cite{middleton}, using a
similar distribution for $T = 0$, suggested an SOPT with a small
order parameter exponent $\beta = 0.017(5)$. Malakis and Fytas \cite{malfyt},
by applying the critical minimum-energy subspace scheme in conjunction
with the Wang–Landau and broad-histogram methods for cubic lattices,
proved that the specific heat and susceptibility are non-self-averaging
using the bimodal distribution.

Another notable manifestation of randomness is the spin glass (SG)
phase exhibited by many systems under certain conditions. These
are random magnetic systems in which the interactions between the
spins are in conflict to each other, a phenomenon known as
frustration, a result of strong frozen-in structural disorder
according to which no single spin configuration is favored by all
interactions, quenched randomness. Moreover, SG is an emergent
phase of matter in random magnetic systems and thus a lot of
studies on disordered systems concern that phase, where the
magnetic and non-magnetic components, making up the material, are
randomly distributed in space; the disorder is present in the
Hamiltonian in the form of random couplings between two
constituent spins, which vary, in general, in their values and
signs according to a PDF $P(J_{ij})$ chosen suitably
\cite{nishimortiz,binderyoung,binder,nishim1,mezardetal,fischer,uspekhi}.
The competing interactions are ferromagnetic and
antiferromagnetic. Conventional SGs are dilute magnetic alloys
such as AuFe or CuMn. The main objective is to understand better,
at a theoretical level, what are the microscopic mechanisms
leading to such a behavior and how to describe them. Many studies,
mainly using the mean-field analysis, have been successful in
elucidating various concepts for understanding SGs. One of the
current issues in SGs is their nature in finite dimensions below
the upper critical dimension. Unfortunately, for finite
dimensions, the calculations often rely on numerical simulations,
because there are few ways to analytically study SGs. Long
equilibration times for their numerical simulations are needed and
average over many realizations of random systems to make error
bars small enough. It is thus difficult to gain a conclusive
understanding on the nature of them in finite dimensions.
Establishing reliable analytical theories of SGs have been one of
the most challenging problems for years. Theoretical physicists
have developed mathematically heuristic tools based on what is
called the ``replica trick''. Another successful analysis to
elucidate their properties is the use of gauge symmetry, by which
one can obtain the exact value of the internal energy, evaluate
the upper bound for the specific heat, and obtain some correlation
inequalities in a subspace known as the Nishimori line
\cite{nishim2}. Since the Nishimori line is also invariant under
renormalization group transformations, the intersection of the
Nishimori line and the FM/PM transition line must be a fixed
point. The so-called Nishimori point corresponds to a new
universality class belonging precisely to the family of strong
disorder fixed points. Moreover, Kaneyoshi has also applied the
effective field theory to the SGs \cite{eft}.

In the parameter plane spanned by temperature and external
magnetic field the high temperature phase is separated from the
spin glass one by the so called de Almeida-Thouless line
(AT-line); consequently, the determination of the AT-line is a
matter of great urgency in the theoretical analysis of the SGs
model. The equilibrium properties of mean field SGs are calculated
by using the two available different approaches. The first one is
the replica method starting with $n$ replicas of the system under
consideration. The free energy can be determined by the saddle
point method. The main feature of the replica method is that the
mathematically problematic limit $n \rightarrow 0$ is usually
taken at the end. In this framework, the AT-line is determined by
the local stability of the replica symmetric saddle point. In the
other method, the cavity one, one spin is added to a system of N
spins and a stochastic stability of the thermodynamic limit $N
\rightarrow \infty$ is used to derive self-consistent equations
for the order parameters. In the latter procedure, the AT-line is
obtained by investigating the correlations between two spins which
vanish in the thermal limit for a pure state of a mean field
system \cite{EA,SK,vanhemmen,wada,parisi,TAP,AT,toninelli}.

An early attempt in the theory of SGs was put forward by Edwards
and Anderson (EA) \cite{EA}, based on Ising model with the
disorder in the exchange integral between nearest neighbors; they
managed to demonstrate the existence of the spin glass phase
within the mean field theory in conjunction with the replica
trick; they also identified two features for a spin glass theory,
frustration and disorder. As their main result was the
introduction of a new type of "order" parameter, which describes
the long-time correlations $q=<<S_{i}^{2}>_{T}>_{J}\neq0$, where
$<…>_{J}$ means configuration averaging over the distributions
$P(J_{ij})$ for all spin pairs $(ij)$ and $<…>_{T}$  means thermal
averaging. A simple mean field approximation leads to $q(T)\neq0$
below a characteristic temperature $T_{f}$ and to a sharp second
order phase transition at $T_{f}$. Their model was a
generalization of the Ising model (Ising spin-glass, ISG) but with
a non constant exchange integral interaction, namely,

\begin{equation}
 H=-\sum_{<i,j>}J_{ij}S_{i}S_{j}  \hspace{2mm},
    \hspace{20mm} S_{i}=\pm1 \label{eahamil}
\end{equation}

where $<i,j>$ implies summation over nearest neighbors, $J_{ij}$
is the bilinear exchange interaction between nearest-neighbor
pairs, randomly quenched variables, identically and independently
distributed according to the single Gaussian probability
distribution function

\begin{equation}
P(J_{ij}) =  [(2 \pi)^{1/2}J]^{-1} \exp\left[-J_{ij}^{2}/(2
J^{2})\right] \label{eagexp}
\end{equation}

with zero mean value and variance $J^{2}$. The disorder is
quenched, in that $J_{ij}$, initially, are chosen randomly but
then fixed for all thermodynamic processes. The ISG together with
the RFIM constitute two of the most-studied subjects in the area
of the disordered magnetic systems. The EA model is far too
difficult to be analyzed theoretically in detail, thus Sherrington
and Kirkpatrick (SK) \cite{SK} introduced in 1975 a simplified
version of this model by replacing the pair interaction by a
long-range one, in that, each particle interacts with the
remaining ones, so that the Hamiltonian (\ref{eahamil}) is
replaced by

\begin{equation}
 H=-\frac{1}{2}\sum_{(ij)}J_{ij}S_{i}S_{j}  \hspace{2mm},
    \hspace{20mm} S_{i}=\pm1 \label{skhamil}
\end{equation}

where $(ij)$ implies summation over all pairs of spins; the
bilinear spin interactions $J_{ij}$ are randomly quenched
variables, specified by a symmetric matrix $\{J_{ij}\}$ and are
distributed according to the probability distribution function

\begin{equation}
P(J_{ij}) =  [(2 \pi)^{1/2}J]^{-1}
\exp\left[-(J_{ij}-J_{0})^{2}/(2 J^{2})\right] \label{skexp}
\end{equation}

with $J$ and $J_{0}$ scaled by,

\begin{equation}
 J=\widetilde{J}/N^{1/2},  J_{0}=\widetilde{J}_{0}/N  \label{scale}
\end{equation}

both $\widetilde{J}$ and $\widetilde{J}_{0}$ are intensive
quantities, so that in the SK model each spin interacts with each
other one via weak interaction of the order $N^{-1/2}$. The
presence of $N$ (the total number of spins in the system) is
necessary to ensure that the respective thermodynamic quantities
are extensive. This simple model captures the basic ingredients of
spin glass physics, namely, quenched randomness and frustration,
and was solved ``exactly'' at the mean field level. The
fluctuations around the SK saddle point are described by an
($\frac{n(n-1)}{2}*\frac{n(n-1)}{2}$) matrix and its eigenvalues
have been determined in Ref. \cite{AT}. The temperature dependence
of these eigenvalues shows that the replica symmetric saddle point
loses its stability at the phase boundary of the SG phase. The SK
model, since its inception, is still alive and attractive
presenting several challenging issues; it presents a continuous
phase transition in which the spin glass phase has the free energy
landscape composed by many almost degenerated thermodynamic states
separated by infinitely high barriers. Its formalism has gone far
beyond the area of disordered magnetic systems being employed in
many other complex systems, like neural networks, optimization
problems as well as in stock markets and in wireless network
communications \cite{wireless}. Many experimental observations
seem to be in good agreement with the predictions of this model.
An important suggestion concerning the behavior of various
physical parameters in the SG phase is the so-called
Parisi-Toulouse (PaT) hypothesis, or projection hypothesis, for
the SK model, according to which the entropy and the EA order
parameter $q$ are field independent $\Bigl(S(T,h)=S(T),
q(T,h)=q(T)\Bigr)$, whereas magnetization is temperature
independent $\Bigl(m(T,h)=m(h)\Bigr)$, where $h$ is an external
magnetic field; Monte Carlo results show strong evidence that
entropy is independent of the applied field,
\cite{partou,vtp,monbou,fischer1,ma}.

In addition to the SK model other infinite-range spin glass models
have been proposed, including Blume-Emery-Capel, Potts, spin-S
($S>1/2$) and vector spin glass models \cite{binderyoung,fischer,uspekhi}.
New phases, characterized by different classes of order
parameters, have emerged, opening many controversial
problems from both theoretical and experimental points of view.

The paper is organized as follows. In the next section, we discuss
the limiting procedure between the canonical partition function
and the Helmholtz free energy, the replica approach and suggest
the joint Gaussian PDF with correlation function $\rho$.  In
section $3$, we introduce the current model with the random field
and calculate the respective free energy functional as well as the
magnetization and the Edwards-Anderson parameter. In section $4$
we present the numerical results, the phase diagram and other
thermodynamic quantities and we close with the conclusions and
discussions in section $5$.

\vspace{-7mm}

\section{The free energy and replica approach}

\noindent

\vspace{-7mm}

As in every problem in equilibrium statistical physics the central
issue is the calculation of the free energy per particle from the
respective one of the N-particle system $F(\beta,N)$ in the
thermodynamic limit, namely

\begin{equation}
 f(\beta) = \lim_{N \rightarrow \infty} \frac{1}{N} F(\beta,N)   \label{fe1}
\end{equation}

where $-\beta F(\beta,N) = \ln Z(\beta,N)$, $Z(\beta,N)$ is the
canonical partition function. However, after the introduction of
randomness, as in the case of spin glasses, its influence on the
system has to be considered, so that the relation (\ref{fe1})
converts into,

\begin{equation}
-\beta f(\beta) = \lim_{N \rightarrow \infty} \frac{1}{N}\Bigl<\ln
Z(\beta,N)\Bigr>_{r}   \label{fe2}
\end{equation}

where $\bigl< \ldots \bigr>_{r}$ represents the thermal average as
well as the one with respect to the randomness, thus the
calculation of the free energy per particle is transferred to the
calculation of $\bigl<\ln Z(\beta,N)\bigr>_{r}$.

The eventual aim is to calculate the various observables, but to
achieve this we calculate initially the Helmholtz free energy $F$;
the calculation of the partition function is a very hard task,
resulting the need for a new procedure to make the calculation
feasible; as far as the function needed is the logarithm of the
partition function and not the partition function itself, the
following formula can be applied,

\begin{equation}
  \ln Z = \lim_{n \rightarrow 0}\frac{Z^{n}-1}{n}   \label{logz}
\end{equation}

implying that we have considered $n$ replicas of the initial
system, which are not interacting, and $Z^{n} = \prod _{\alpha
=1}^{n} Z_{\alpha}$, where $\alpha$ is the replica identifier and
instead of averaging $\ln Z$ we average $Z^{n}$. Considering this
expression for $\bigl<\ln Z \bigr>$, the relation (\ref{fe2})
converts into

\begin{equation}
 -\beta f(\beta) =  \lim_{n \rightarrow 0} \lim_{N \rightarrow \infty}
 \frac{1}{Nn}\left( \Bigl< Z^{n}\Bigr>_{r}-1\right)   \label{fe3}
\end{equation}

the order of the limits is irrelevant \cite{vanhemmen}, although
the thermodynamic one precedes that of $n \rightarrow 0$ in order
to apply the steepest descent method; the replicated partition
function for integer values of $n$ assumes the form

\begin{equation}
Z^{n}(\beta) = \sum _{\{S^{\alpha}_{i}=\pm 1\}} \exp \Big[ -\beta
\sum_{\alpha=1}^{n} H \Big( \{ S^{\alpha} _{i} \} \Big) \Big]
\label{partitionfn}
\end{equation}

Sherrington and Kirkpatrick  managed to derive an expression for
the respective free energy, by calculating initially the free
energy functional with respect to the two parameters $m_{\alpha}$
and $q_{\alpha \gamma}$ dependent on replicas, introduced through
the Hubbard-Stratonovich transformation, with $\alpha$ and
$\gamma$ characterizing the replicas and $\alpha \neq \gamma$,
among other parameters; by considering the replica symmetry
hypothesis (RS), that is $m_{\alpha}=m$ and $q_{\alpha \gamma}=q$
for every $\alpha$ and $\gamma$, and using the analytic
continuation $n \rightarrow 0$ succeeded in calculating the system
free energy, namely

\begin{equation}
\beta F  = N \left\{
-\frac{\widetilde{J}^{2}\beta^{2}}{4}(1-q)^2+\frac{\widetilde{J}_{0}m^{2}\beta}{2}
  -\frac{1}{(2\pi)^{1/2}} \int dz e^{-z^{2}/2}\ln\left[2\cosh(\widetilde{H}(z)) \right] \right\}   \label{SKfe}
\end{equation}

where $\widetilde{H}(z) = \beta \widetilde{J}q^{1/2}z + \beta
\widetilde{J}_{0}m$, $m=<<S_{i}>_{T}>_{J}$ and
$q=<<S_{i}>_{T}^{2}>_{J}$, in conjunction with the exponential
identity (Hubbard-Stratonovich transformation)

\begin{equation}
  e^{\frac{\lambda \sigma^{2}}{2}}=\Biggl(\frac{\lambda}{2 \pi}\Biggr)^{1/2}
    \int_{-\infty}^{\infty} e^{-\frac{\lambda x^{2}}{2}+\lambda \sigma x} dx  \label{hstr}
\end{equation}

The SK model was solved by means of the replica method and,
consequently, it was originally thought that it was exactly
solvable, although Sherrington and Kirkpatrick were aware that it
suffers from a serious drawback, in that, it possesses a negative
entropy at zero temperature, specifically
$S(T=0^{\circ}K)=-Nk/2\pi$. SK model was the subject of a large
number of publications and cannot be considered as a trivial
one in the mean field sense. Though rather unrealistic, it seems
to describe some spin glass properties correctly. The SK model for
Ising spins with quenched random bonds is the simplest
representative of a class of long-ranged models all successfully
describing the interesting phenomena of spin glasses. In addition
to this success in physical questions, the research on these
models has been fruitful and stimulating in optimization problems,
in understanding the neural networks and communications. The
zero-temperature entropy anomaly ($s(T=0)<0$), appearing in the SK
model, has been shown to be associated with the hypothesis of the
replica symmetry of the two order parameters, $m$ and $q$. The
correct low-temperature solution, resulting by breaking the
replica symmetry of SK, was proposed by Parisi \cite{parisi}, and
consists of a continuous order parameter function (an infinite
number of order parameters) associated with many low-energy
states, a procedure which is usually called the replica-symmetry
breaking (RSB). Parisi was the first who found a satisfactory
solution of the SK model in the SG regime. Curiously, the simplest
one-step RSB (1S-RSB) procedure improves, in part, this anomaly,
in the sense that the zero temperature entropy per particle
becomes less negative, from $s(0)/(Nk_{B})\simeq-0.16$ within the
RS it rises to $s(0)/(Nk_{B})\simeq-0.01$ within the 1S-RSB, a
significant improvement. The complementary approach by Thouless,
Anderson and Palmer (TAP), for investigating the spin glass model,
does not perform the bond average, and permits a treatment of
problems depending on specific configurations \cite{TAP}. For
other questions which are expected to be independent of the
special configuration, such as all macroscopic physical
quantities, self averaging occurs. This is due to the fact that
the random interaction matrices have well-known asymptotic
properties in the thermodynamic limit. The situation is in
principle similar to the central limit theorem in probability
theory, where large numbers of random variables also permit the
calculation of macroscopic quantities which hold for nearly every
realization of the random variables. Thus the investigation of one
or some representative systems is sufficient and the bond average
is not needed. The TAP equations have been well established for
more than two decades and several alternative derivations are
known. Nevertheless the TAP approach is still a field of current
interest. This is due to the importance of the approach to
numerous interesting problems. Moreover it is suspected that not
all aspects of this approach have yet been worked out.
Furthermore, a transition in the presence of an external magnetic
field, known as the Almeida-Thouless (AT) line
\cite{AT,toninelli}, is found in the solution of the SK model:
such a line separates a low-temperature region, characterized by
RSB, from a high-temperature one, where a simple one-parameter
solution, RS solution, is stable. Numerical simulations are very
hard to carry out for short-range ISG's on a cubic lattice, due to
large thermalization times; as a consequence, no conclusive
results in three-dimensional systems are available. However, in
four dimensions the critical temperature is much higher, making
thermalization easier; in this case, many works claim to have
observed some mean-field features.

Reentrant spin-glass (RSG) transition is a well-known phenomenon
of spin glasses. The RSG transition is found near the phase
boundary between the SG phase and the FM phase. As the temperature
decreases from a higher temperature, magnetization once increases
and then disappears at a lower temperature. Finally, the SG phase
is realized. The phenomenon was first considered as a phase
transition between an ''FM phase'' and a ''SG phase''. However,
neutron diffraction studies have revealed that the ''SG phase'' is
characterized by FM clusters. Now the RSG transition is believed
to be a reentry from a FM phase to a frozen state with FM
clusters. The mechanism responsible for this reentrant transition
has not yet been resolved. Two ideas have been proposed for
describing the RSG transition: (i) an infinite-range Ising bond
model, and (ii) a phenomenological random field concept. The
essential point of that conception is that the system is
decomposed into an FM part and a part with frustrated spins (SG
part). At low temperatures, the spins of the SG part yield random
effective fields to the spins of the FM part. Nevertheless, no
theoretical evidence has yet been presented for this idea in a
microscopic point of view. In the last two decades, computer
simulations have been performed extensively to solve the RSG
transition in various models such as short-range bond models
\cite{fischer,physicstoday,imryma}, short-range site models
\cite{aharony,fishaha}, and a Ruderman-Kittel-Kasuya-Yoshida model
\cite{imbrie,galam3}.

Although the aforementioned types of randomness in magnetic
systems consist a significant branch of statistical physics, very
few investigations have considered them together
\cite{nombreetal}, but even then the considered probability
density function for the random bonds and random fields were
considered as distinct and their joint probability density
function was simply their product so that the one type of
randomness does not influence the other one directly and in fact
these are two independent random variables. There are systems
described by ISG in the presence of random fields, such as proton
and deuteron glasses, mixtures of hydrogen-bonded ferroelectric
and antiferroelectrics \cite{fer,kim}. The diluted
antiferromagnets, such as $Fe_{x}Zn_{1-x}F_{2}$, under the
influence of a uniform magnetic field form the realization of an
RFIM \cite{young}. For $x\leq 0.24$ it becomes an ISG, whereas for
$x \geq 0.40$ it is an RFIM. For $0.24\leq x \leq 0.40$ both
behaviors appear, RFIM or ISG for small or large magnetic fields,
respectively. Also, the compound $CdCr_{1.7}Ir_{0.3}S{4}$ in a
magnetic field exhibits all the characteristic features of SGs
\cite{lefloch} as well as $LiHo_{x}Er_{1-x}F_{4}$ for various
values of $x$ \cite{piatek}.

In the current investigation this restriction concerning the
discreteness of the PDFs is lifted by considering a pure joint
Gaussian probability density function for $J_{ij}$ and $h_{i}$ as

\vspace{-7mm}

\begin{eqnarray}
 P(J_{ij},h_{i}) & = & \frac{N^{1/2}}{2\pi \Delta J (1-\rho^{2})^{1/2}}\
 \exp \Biggl\{ -\frac{1}{2(1-\rho^{2})} \Biggl[ N\frac{(J_{ij}-J_{0}/N)^{2}}
 {J^{2}} - \nonumber  \\
   &  &  2\rho N^{1/2} \frac{(J_{ij}-J_{0}/N)(h_{i}-h_{0})}{\Delta J}
   +\frac{(h_{i}-h_{0})^{2}}{\Delta^{2}}  \Biggr]      \Biggr\}   \label{joint}
\end{eqnarray}

\vspace{-5mm}

where $\rho$ is the correlation function (or, simply, correlation)
of the two random variables $J_{ij},h_{i}$ with $\rho =
Cov(J_{ij},h_{i})/(J \Delta)$, $Cov(J_{ij},h_{i})$ their
covariance; $h_{0}, \Delta^{2}$ are the mean value of the random
field and its variance, respectively. This joint PDF shall be used
for the study of the SK spin glass model in the presence of a
random field.

\vspace{-7mm}

\section{The Sherrington-Kirkpatrick spin glass model in the
presence of a random field - Replica formalism}
\noindent

\vspace{-7mm}

The Sherrington-Kirkpatrick infinite-range model of spin glasses
for Ising spins $S_{i}=\pm 1, i=1,2,...,N$, with random pair
interactions (specified by a symmetric matrix $\{J_{ij} \}$) in
the presence of a random field $\{h_{i} \}$ is described by the
Hamiltonian

\begin{equation}
  H=- \frac{1}{2}\sum_{(i,j)}J_{ij}S_{i}S_{j} -\sum_{i=1}^{N} h_{i}S_{i}    \label{hamskrf}
\end{equation}


where the first sum runs over all pairs of spins and indicated by
$(i,j)$. The exchange interactions $\{J_{ij}\}$ and random fields
$\{h_{i}\}$ are quenched random variables drawn from the joint
Gaussian PDF (\ref{joint}). The analysis of this model shall be
relied on the MFA as in Ref. \cite{SK}; although this direction of
study, in the most cases, for ordinary systems (i.e.,
ferromagnetic, etc.) is very simple and practically trivial, this
does not happen to be for the SG case; for the latter case this
study is highly no trivial even for the simplest SG, the ISG, thus
motivating intensive research activity in this direction, deriving
a plethora of publications and new ideas.

Considering, now, a realization of the bonds and random fields $(\{J_{ij}\},\{h_{i}\})$,
the respective free energy $F(\{J_{ij}\},\{h_{i}\})$ results as an average
over both disorders, in addition to the thermal one,

\begin{equation}
 \Biggl< F(\{J_{ij}\},\{h_{i}\}) \Biggr>_{J,h} = \int \prod _{(ij)}\prod_{i} P(J_{ij},h_{i})
    F\Bigl(\{J_{ij}\},\{h_{i}\}\Bigr) dJ_{ij} dh_{i}      \label{jfren}
\end{equation}

so that the free energy per particle $f$ assumes the form with
respect to the partition function,

\begin{equation}
 -\beta f(\beta) =  \lim_{n \rightarrow 0} \lim_{N \rightarrow \infty}
 \frac{1}{Nn}\left( \Bigl< Z^{n}\Bigr>_{J,h}-1\right)   \label{rfe1}
\end{equation}

Within the framework of the replica method, the average partition
function $ \Bigl< Z^{n}\Bigr>_{J,h}$ over both disorders for
integer $n$ is calculated to be

\begin{eqnarray}
\Bigl< Z^{n}\Bigr>_{J,h} = \sum_{\{ S_{\alpha}=\pm 1 \}} \exp
\Biggl\{ \frac{(\beta \Delta)^{2}}{2} \sum_{i=1}^{N} \Bigl(
\sum_{\alpha=1}^{n} S_{i}^{\alpha} \Bigr)^{2} + \frac{(\beta
J)^{2}}{2N} \sum_{(i,j)} \Bigl( \sum_{\alpha=1}^{n}
S_{i}^{\alpha} S_{j}^{\alpha} \Bigr)^{2} +                      \nonumber   \\
\frac{\beta^{2} \rho J \Delta}{\sqrt{N}} \sum_{i=1}^{N} \Bigl(
\sum_{\alpha=1}^{n} S_{i}^{\alpha} \Bigr) \sum_{(k,l)} \Bigl(
\sum_{\alpha=1}^{n} S_{k}^{\alpha} S_{l}^{\alpha} \Bigr) +
 \beta h_{0} \sum_{i=1}^{N} \Bigl(
\sum_{\alpha=1}^{n} S_{i}^{\alpha} \Bigr) + \frac{\beta J_{0}}{N}
\sum_{(i,j)} \Bigl( \sum_{\alpha=1}^{n} S_{i}^{\alpha}
S_{j}^{\alpha} \Bigr)  \Biggr\}     \;\;\;\;\;\  \label{partfn1}
\end{eqnarray}

where $\alpha$ characterizes the replica and $i$ the lattice site.
The above expression can be linearized in the spins by introducing
the replica matrix $\{q_{\alpha \beta}\}$ as well as the auxiliary
quantity $\{m_{\alpha}\}$, then by reordering and dropping out
terms that disappear in the thermodynamic limit
($N\rightarrow\infty$) by using the so-called Hubbard-Stratonovich
transformation (\ref{hstr}) (exponential transformation) one finds

\begin{eqnarray}
\Bigl< Z^{n}\Bigr>_{J,h} = \exp \Biggl\{ \frac{\beta^{2}
\Delta^{2} n N}{2} + \frac{\beta^{2}J^{2}n N}{4}\Biggr\}
\int_{-\infty}^{\infty} \Biggl( \prod_{\alpha \gamma}
\Biggl(\frac{\beta^{2}J^{2}N}{2\pi} \Biggr)^{1/2}dq_{\alpha
\gamma} \Biggr) \Biggl( \prod_{\alpha} \Biggl(\frac{\beta
J_{0}N}{2\pi}\Biggr)^{1/2}dm_{\alpha} \Biggr)                               \nonumber   \\
 \Biggl( \prod_{\kappa} \Biggl(\frac{\nu N}{2\pi}
\Biggr)^{1/2}dm_{\kappa}\Biggr) \Biggl( \prod_{\delta}
\Biggl(\frac{\nu N}{2\pi} \Biggr)^{1/2}dm_{\delta}\Biggr) \Biggl(
\prod_{\lambda} \Biggl(\frac{\nu N}{\pi} \Biggr)^{1/2}dm_{\lambda}\Biggr)        \nonumber   \\
 \Biggl( \prod_{\varepsilon} \Biggl(\frac{\nu N}{2\pi}
\Biggr)^{1/2}dm_{\varepsilon}\Biggr) \int_{-i\infty}^{i\infty}
\Biggl( \prod_{\eta} \Biggl(\frac{i \nu N}{\pi} \Biggr)^{1/2}dm_{\eta}\Biggr)             \nonumber   \\
 \exp \Biggl\{ -N \Biggl[ \frac{(\beta J)^{2}}{4} \sum _{(\alpha \gamma)} q_{\alpha \gamma}^{2}+
\frac{\beta J_{0}}{2} \sum _{\alpha=1}^{n} m_{\alpha}^{2} +
\frac{\nu}{2} \sum _{\kappa=1}^{n} m_{\kappa}^{2} + \nu \sum
_{\delta=1}^{n} m_{\delta} \sum _{\lambda=1}^{n} m_{\lambda}^{2} +  \nonumber   \\
-i \nu \sum _{\varepsilon=1}^{n} m_{\varepsilon} \sum
_{\eta=1}^{n} m_{\eta}^{2} \Biggr] \Biggr\} \;\;
Tr_{\{S^{\alpha}\}} \exp \Bigl[ N Z_{n}\Bigl( \{q_{\alpha
\gamma}\}, \{m_{\alpha}\} \Bigr) \Bigr]    \;\;\;\;\;\;\;\
\label{partfn2}
\end{eqnarray}

where $\nu=\beta^{2} \rho \Delta J/2$ and

\begin{eqnarray}
Z_{n}\Bigl( \{q_{\alpha \gamma}\},\{m_{\alpha}\} \Bigr) =
\frac{(\beta \Delta)^{2}}{2}\sum_{(\alpha,\gamma)}
S^{\alpha}S^{\gamma} + \beta h_{0} \sum_{\alpha=1}^{n} S^{\alpha}+
\frac{(\beta J)^{2}}{2}  \sum_{(\alpha
\gamma)}q_{\alpha \gamma}S^{\alpha}S^{\gamma} +                    \nonumber   \\
\beta J_{0} \sum_{\alpha=1}^{n} m_{\alpha}S^{\alpha}+ \nu
\sum_{\kappa=1}^{n} m_{\kappa}S^{\kappa} - i\nu
\sum_{\varepsilon=1}^{n} m_{\varepsilon}S^{\varepsilon}
\;\;\;\;\;\;   \label{partfn3}
\end{eqnarray}

The respective integrals in (\ref{partfn2}) were evaluated by the
steepest descent method, since the exponential argument is
proportional to $N$. In expressions (\ref{partfn2}) and
(\ref{partfn3}), the site indices ($i$ or $j$) are absent on the
spins since all of the sites are equivalent, thus the resulting
$N$ can be extracted out simplifying the calculation of the
thermodynamic limit ($N \rightarrow \infty$) and ($\alpha \gamma$)
indicates a distinct pair of replicas with $\alpha \neq \gamma$.
The trace in (\ref{partfn2}) is over the $n$ replicas at a single
spin site. The two limits, the thermodynamic one ($N \rightarrow
\infty$) and the analytic continuation ($n \rightarrow 0$), can be
interchanged \cite{vanhemmen}; in order to calculate the free
energy per particle from the expressions (\ref{rfe1}),
(\ref{partfn1}), (\ref{partfn2}) and (\ref{partfn3}), the two
limits are performed consecutively; first we consider the
thermodynamic limit yielding

\begin{eqnarray}
-\beta f(\beta) = \left(\frac{\beta J}{2}\right)^{2}+ \frac{(\beta
\Delta)^{2}}{2} -                                                   \nonumber   \\
\lim_{n \rightarrow 0}\frac{1}{n} \left[ \left(\frac{\beta
J}{2}\right)^{2} \sum_{(\alpha \gamma)} q^{2}_{\alpha \gamma}
+\left(\frac{\beta J_{0}+\nu}{2}\right)
\sum_{\alpha}m^{2}_{\alpha} + (1-i) \nu \sum_{\alpha} m_{\alpha}
\sum_{\gamma} m_{\gamma}^{2} \right] +                              \nonumber   \\
\lim_{n \rightarrow 0} \frac{1}{n} \ln Tr \exp \Biggl\{
\frac{(\beta \Delta)^{2}}{2} \sum_{(\alpha \gamma)} S^{\alpha}
S^{\gamma} +\beta h_{0} \sum_{\alpha}S^{\alpha} + \frac{(\beta
J)^{2}}{2} \sum_{(\alpha \gamma)} S^{\alpha} S^{\gamma} q_{\alpha
\gamma} +                                                          \nonumber   \\
 (\beta J_{0} +\nu) \sum_{\alpha} m_{\alpha}S^{\alpha}
 - i\nu \sum_{\alpha}m_{\alpha} S^{\alpha} \Biggr\}
 \;\;\;\;\;\;\;\;\;\;\;\   \label{rfe2}
\end{eqnarray}

In order to perform the analytic continuation $n \rightarrow 0$,
we invoke the RS hypothesis, namely, $q_{\alpha \gamma}=q$ and
$m_{\alpha}=m$ for every $\alpha$ and $\gamma$, then taking the
limit $n \rightarrow 0$ the free energy in (\ref{rfe2}) assumes
the form, in conjunction with the Hubbard-Stratonovich
transformation (\ref{hstr})

\begin{eqnarray}
\beta f(\beta) = -\left(\frac{\beta J}{2}\right)^{2}(1-q)^{2}
+\frac{2\beta J_{0} + \beta^{2} \rho \Delta J}{4}m^{2} - \nonumber \\
   \frac{1}{\sqrt{2\pi}} \int_{-\infty}^{\infty} dz\;e^{-z^{2}/2}
   \ln\Biggl\{2\Bigl[\cosh(A(z)) \cos(B) + i \sinh(A(z)) \sin(B) \Bigr] \Biggr\}  \label{rfe3}
\end{eqnarray}

where $A(z)= \beta \left[h_{0} + m\left(J_{0}+
\frac{1}{2}\beta\rho\Delta J\right) + z (qJ^{2} +
\Delta^{2})^{1/2}\right] $ and $B = -\beta^{2} \rho \Delta J m/2$.
The expression for $A(z)$ and $B$ contain the interdependence of
both disorders. Eq. (\ref{rfe3}) converts into the one found by
Sherrington and Kirkpatrick \cite{SK} in the absence of random
fields, i.e., for $h_{0}=0,\; \Delta=0$. The free energy for the
system resulting from (\ref{rfe3}) as a function of the
temperature and the PDF's parameters is

\begin{eqnarray}
f(\beta) = -\frac{\beta J^{2}}{4} (1-q)^{2} +\frac{2J_{0} + \beta
\rho \Delta J}{4} \; m^{2} -  \nonumber   \\
  \frac{kT}{2\sqrt{2\pi}} \int_{-\infty}^{\infty} dz\;e^{-z^{2}/2}
\ln\left[ 4 \left(\cosh^{2}A - \sin^{2}B\right) \right]
\label{rfe4}
\end{eqnarray}

The former expression constitutes the principal quantity of the
current investigation. The logarithmic function in (\ref{rfe4}) is
not singular, since the functions $\cosh(A(z))$ and $\sin(B)$ do
not intersect each other, so the integrand is well-defined. The
quantities $q, m$ are given self-consistently by the extremum
conditions (saddle-point conditions), $\frac{\partial(\beta
f)}{\partial q}=0$ and $\frac{\partial(\beta f)}{\partial m}=0$
from (\ref{rfe4}), so that the two principal quantities $m$ and
$q$ satisfy the simultaneous equations

\begin{eqnarray}
m & = & \frac{1} {2 J_{0} + \beta \rho \Delta J}
\frac{1}{2\sqrt{2\pi}}
  \int_{-\infty}^{\infty} e^{-z^{2}/2} \frac{\left(2J_{0} + \beta \rho \Delta J \right)
  \sinh(2A(z)) + \beta \rho \Delta J \sin(2B)} {\cosh^{2}(A(z)) - \sin^{2}B}\;dz       \nonumber \\
q & = & \frac{1}{4\sqrt{2\pi}} \int_{-\infty}^{\infty}
e^{-z^{2}/2}    \frac{\sinh^{2}(2A(z)) - \sin^{2}(2B)}{\left(
\cosh^{2}(A(z)) - \sin^{2}B \right)^{2}} \; dz    \label{mqequs}
\end{eqnarray}

which become identical to those found by Sherrington and
Kirkpatrick \cite{SK} in case the random fields are absent. Both
denominators in (\ref{mqequs}) do not vanish for the same reason
as in a previous paragraph.

Initially, we shall focus our attention on the so called
Parisi-Toulouse (PaT) hypothesis to explore what occurs in the low
temperature SG phase
\cite{binderyoung,partou,vtp,monbou,fischer1,ma}; consequently,
the integrand in (\ref{rfe4}) is expanded for small values of the
arguments (small $h_{0}$),
\cite{binderyoung,partou,vtp,monbou,fischer1,ma}, namely

\begin{eqnarray}
-\beta f  =  \frac{(\beta J)^2}{4} (1-q)^2 - \beta \frac{2J_{0}
+ \beta \rho \Delta J}{4}m^2 + \ln 2 +                               \nonumber \\
 \frac{1}{2} \left[2\ln \cos B+ \frac{\alpha_{1}^{2}+\alpha_{2}^{2}}{\cos^{2}B}
-\frac{2\sin^{2}B +1}{6\cos^{4}B}
(\alpha_{1}^{4}+3\alpha_{2}^{4}+6\alpha_{1}^{2}\alpha_{2}^{2})
\right] \label{fenew}
\end{eqnarray}

where $\alpha_{1}=\beta (h_{0} + (2J_{0}+\beta \rho \Delta J)m/2)$,
$\alpha_{2} = \beta (\Delta^{2} +qJ^{2})^{1/2}$. Considering the
extremum $\frac{\partial(-\beta f)}{\partial q} = 0$ we get

\begin{equation}
q = \frac{(1+2\sin^{2}B)(\alpha_{1}^{2} + \beta ^{2} \Delta
^{2})-\sin^{2}(2B)/4 } {\cos^{4}B - (1+2\sin^{2}B)\beta^{2}J^{2}}
           \label{qnew1}
\end{equation}

the respective equation for the magnetization $m$, resulting from
the extremum condition $\frac{\partial(-\beta f)}{\partial m} =
0$, is

\begin{eqnarray}
(\beta J_{0}+w)m = -w \tan(v/2) + \frac{\alpha_{1}(\beta
J_{0}+w)}{\cos^{2}(v/2)} + \frac{w\sin(v)}{2}\frac{\alpha^{2}_{1}
+ \alpha^{2}_{2}}{\cos^{4}(v/2)}                  \nonumber \\
- \frac{w\sin(v)}{6} \; \frac{2 + \sin^{2}B}{\cos^{6}(v/2)}
(\alpha_{1}^{4} + 3 \alpha_{2}^{4} +6\alpha_{1}^{2}
\alpha_{2}^{2})                                    \nonumber \\
- \frac{\beta J_{0}+w}{3} \; \frac{1 +
2\sin^{2}B}{\cos^{4}(v/2)}(\alpha_{1}^{3} +3\alpha_{1}
\alpha_{2}^{2})                                       \label{mall1}
\end{eqnarray}

where $w=\beta^{2} \rho \Delta/2$ and $v=\beta^{2} \rho \Delta m$.
However, since the explicit formulae in (\ref{qnew1}) and (\ref{mall1})
seem to be too lengthy, we resort to consider the special case with
$\rho=0$ (constraint); in this case Eqs. (\ref{qnew1}) and (\ref{mall1}) become

\begin{equation}
q = \beta^{2} \frac{(h_{0}+mJ_{0})^{2} + \Delta ^{2}} {1 - \beta^{2}J^{2}}
           \label{qnew2}
\end{equation}

\begin{equation}
 m = \beta (h_{0}+mJ_{0}) \Biggl\{ 1-\frac{\beta ^{2}}{3} \left[ (h_{0} + mJ_{0})^{2}
 +3\left(\Delta^{2} + qJ^{2}\right) \right] \Biggr\}                \label{mnew2}
\end{equation}

Introducing Eq. (\ref{qnew2}) into (\ref{mnew2}), yields

\begin{equation}
 m = \beta (h_{0}+mJ_{0}) \Biggl\{ 1-\frac{\beta ^{2}}{3} \; \left[ (h_{0} + mJ_{0})^{2}
 +3\left(\Delta^{2} + \beta^{2}J^{2} \frac{(h_{0} + mJ_{0})^{2}+\Delta^{2}}{1-\beta^{2}J^{2}}
 \right) \right] \Biggr\}                \label{mnew2a}
\end{equation}

or

\begin{equation}
 m = \beta (h_{0}+mJ_{0}) \Biggl\{ 1-\frac{\beta ^{2}}{3} \;  \frac{(h_{0} + mJ_{0})^{2} +
 3\Delta^{2} + 2\beta^{2}J^{2}(h_{0} + mJ_{0})^{2}}{1-\beta^{2}J^{2}} \Biggr\}                \label{mnew2b}
\end{equation}

In the absence of an external magnetic field and for $J_{0}=0$ the
transition temperature to the spin glass phase is $T_{f} = J$ (see
\cite{binderyoung}), introducing this into (\ref{mnew2b}) yields

\begin{equation}
 m = \frac{h_{0}+mJ_{0}}{T} \Biggl\{ 1 - \frac{1}{3T^{2}}
 \frac{(h_{0} + mJ_{0})^{2} \left(T^{2}+2 \, T^{2}_{f}\right) + 3\Delta^{2}T^{2}}{T^{2}-T_{f}^{2}}  \Biggr\}
     \;\;\;\;\;\;     T > T_{f} \;\;\;\;       \label{mnew2c}
\end{equation}

implying that magnetization $m$ is a nonanalytic function of
$h_{0}$ at $T = T_{f}$. The respective equation for $q$ is

\begin{equation}
 q = \frac{(h_{0}+mJ_{0})^{2} + \Delta^{2}} {T^{2} - T_{f}^{2}}
 \;\;\;\;\;\;   T>T_{f}    \label{qnewc}
\end{equation}

In addition, for $T = T_{f}$, we find by expanding both Eqs.
(\ref{mqequs})

\begin{eqnarray}
 m & = & \frac{h_{0}}{T_{f}} \Biggl\{ 1+\frac{2\, h_{0}^{2}}{3\,T_{f}^{2}} -
 \frac{h_{0}}{T_{f}\sqrt{2}} \left[ 1 +\frac{2\, h_{0}^{2}}{3\,T_{f}^{2}} +
 \frac{\Delta^{2}}{2\,h_{0}^{2}} \right]  \Biggr\} \nonumber \\
 q & = & \frac{h_{0}}{T_{f}\sqrt{2}}\left[ 1+\frac{2\, h_{0}^{2}}{3\,T_{f}^{2}} +
 \frac{\Delta^{2}}{2\,h_{0}^{2}}  \right] -\frac{h_{0}^{2} +
 \Delta^{2}}{T_{f}^{2}}   \;\;\;\;\;\;\;\;\;\;T = T_{f} \;\;\;\;\; \label{mqtf}
\end{eqnarray}

Eqs. (\ref{mnew2c}), (\ref{qnewc}), (\ref{mqtf}) constitute a
generalization of the so called Parisi-Toulouse (PaT) hypothesis
(or projection hypothesis) in the presence of an external random
magnetic field and for $\Delta = 0$ both equations convert into
the respective ones in Refs.
\cite{binderyoung,partou,vtp,monbou,fischer1}. However, for
$T<T_{f}$ the Parisi-Toulouse hypothesis,
\cite{partou,vtp,monbou,fischer1}, guesses that

\begin{equation}
\frac{m}{h_{0}} = 1 - \left(\frac{3}{4}\right)^{2/3} h_{0}^{4/3} +\frac{7}{6} h_{0}^{2}
                   \;\;\;\;\;\;T < T_{f}  \label{mtstf}
\end{equation}

\vspace{-7mm}

\section{Numerical results. Phase diagrams}

\vspace{-7mm}

In order to simplify the calculations to follow we use as measure
the standard deviation $J$ of the exchange integral in Eq.
(\ref{joint}) by setting from now on $J=1$.

Taking the derivative with respect to $m$ of the first relation in
(\ref{mqequs}) and softening to zero the magnetization $m$, we
find

\begin{equation}
 \frac{2J_{0}T^{2}+\rho\Delta T }{2J_{0} (J_{0} T + \rho\Delta )}=
 \frac{1}{\sqrt{2 \pi}}\int_{-\infty}^{\infty} \frac{e^{-z^{2}/2}}{\cosh^{2}H(z)} \;dz        \label{mboundary1}
 \end{equation}

where $H(z) = \beta \left[ h_{0} + z\left( q + \Delta^{2}
\right)^{1/2} \right]$, from which the value for $J_{0}$ is
calculated, since the temperature $T$ is known a priori,

\begin{eqnarray}
J_{0} & = & \frac{1}{2} \left[ \frac{T}{1-q}-\frac{\rho \Delta}{T} +
\sqrt{\left(\frac{T}{1-q} \right)^{2} +
\left(\frac{\rho \Delta}{T} \right)^{2} } \right] \nonumber   \\
& = & \frac{1}{2} \left[ J_{0}^{(SK)}-\frac{\rho \Delta}{T} +
\sqrt{\left(J_{0}^{(SK)} \right)^{2} + \left(\frac{\rho \Delta}{T} \right)^{2} } \right] \label{mboundary2}
\end{eqnarray}

the positive root is chosen, since the negative one leads to a
negative $J_{0}$ and $J_{0} > J_{0}^{(SK)}$. For
$h_{0}=\Delta=\rho = 0$, the Sherrington-Kirkpatrick result is
recovered, $J_{0}^{(SK)} = \frac{T}{(1-q)}$. From this equation
one can calculate analytically its high as well as low temperature
behavior, including that for zero temperature ($T=0^{\circ}K$).

\begin{figure}[htbp]
\begin{center}
\includegraphics*[height=0.20\textheight]{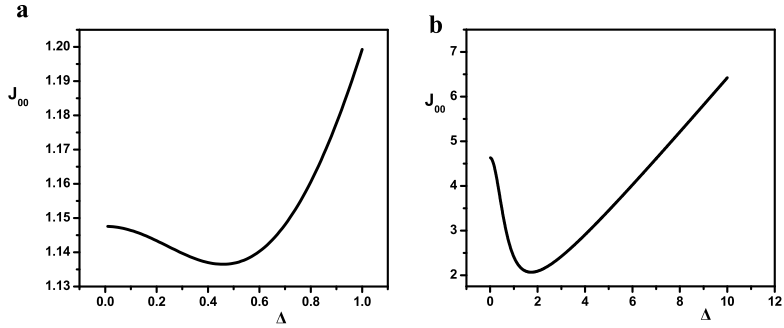}
\caption{\label{figa} The average value $J_{0}$ for $T=0K$
($J_{00}$) as a function of $\Delta$ for specific $h_{0}$-values,
(a) $h_{0}=1.1$ and (b) $h_{0}=2.0$. $J_{00}$ possesses a minimum
value only when $h_{0} \geq 1.0$, since $\Delta = \sqrt{h_{0}^{2}
- 1}$}
\end{center}
\end{figure}

For the zero temperature, one finds

\begin{equation}
J_{0}(T=0)\equiv J_{00} = \sqrt{\frac{\pi}{2}} \sqrt{1+\Delta^{2}}
\exp\left[\frac{h_{0}^{2}}{2(1+\Delta^{2})}\right]
\label{tempzero}
\end{equation}

which, considered as a function of $\Delta$, for a specific
$h_{0}$-value, possesses a minimum value at $\Delta =
\sqrt{h_{0}^{2} - 1}$; this minimum is present only for $h_{0}\geq
1$ and its plots appear in Fig.~\ref{figa}. The low temperature
region one has

\begin{equation}
  \frac{1}{J_{0}} = \frac{1}{J_{00}} + \frac{2T}{\pi
  (1+\Delta^{2})^{2}}                                \label{templow}
\end{equation}

In order to facilitate the evaluation of the phase diagram (the
temperature $T$ with respect to $J_{0}$) initially we focus on the
boundary between the PM ($m=0, q=0$) and SG ($m=0, q\neq0$)
phases, which is achieved by expanding the free energy
(\ref{rfe4}) in powers of $q$ under the constraint $m=0$
\cite{binderyoung,suzuki,nambu1,nambu2}; in this expansion the
coefficient of $q^{2}$-term is set equal to zero in order to
determine the respective transition temperature $T_{f}$ between
the aforementioned phases, yielding

\begin{equation}
 T_{f} = \Bigg\{ \frac{1 \pm \Big[1 -16 (\Delta^{2} +
 h_{0}^{2})\Big]^{1/2}}{2}\Bigg\}^{1/2}              \label{pmsgt}
\end{equation}

being $J_{0}$-independent, $\frac{dT_{f}}{dJ_{0}} = 0$, thus
representing a straight line in ($J_{0}-T$)-plane. However, as it
is clear from (\ref{pmsgt}) the transition temperature $T_{f}$
possesses two branches, the plus-one and the minus-one; both
temperatures lead to a spin glass phase, but in case of absence of
an external magnetic field ($h_{0}=0$ and $\Delta=0$) the
resulting temperature from the plus-branch is $T_{f}^{+}=1$ (in
units of $J$) whereas from the minus-branch is $T_{f}^{-}=0$, that
is, both specify the two limits for the existence of the SG phase;
the former one leads to the existence of a non zero temperature
spin glass phase and not to a trivial temperature as the latter
yields. However, the functional form of Eq. (\ref{pmsgt}) imposes
a severe constraint on the specific values $\Delta$ and $h_{0}$
can assume for the existence of a spin glass phase, since they
have to satisfy the relation $\Delta^{2} + h_{0}^{2} \leq 0.0625$
so that the interior square root is meaningful. Due to this
constraint, the temperature $T_{f}^{+}$ possesses a minimum value,
which is $T_{f}^{+} = \sqrt{0.5}$, which is the maximum one for
the $T_{f}^{-}$, whose minimum value is zero, $T_{f}^{-} = 0.0$.

\begin{figure}[htbp]
\begin{center}
\includegraphics*[height=0.20\textheight]{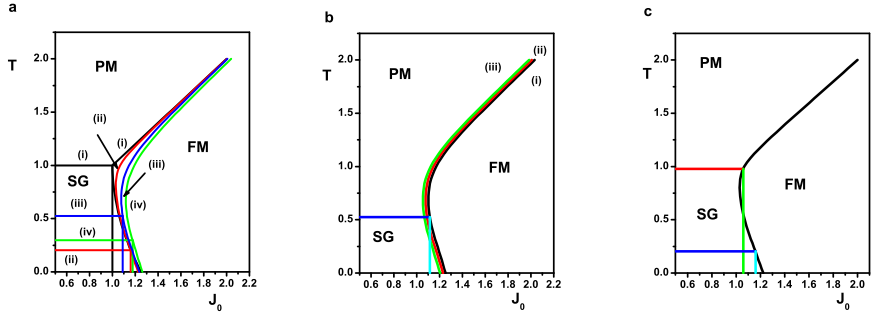}
\caption{\label{figb} (Color online) (a) The phase diagram in the
absence of a random field (i) and in presence of a random field
(ii) $\Delta=0.1, h_{0}=0.0, \rho=0.1$, (iii)  $\Delta=0.1,
h_{0}=0.1, \rho=0.1$, (iv)  $\Delta=0.1, h_{0}=0.2, \rho=0.5.$ (b)
The phase diagram for fixed $\Delta=0.1, h_{0}=0.2$ and (i)
$\rho=0.0$, (ii) $\rho=0.5$ and (iii) $\rho=1.0$; as the
correlation $\rho$ increases the extent of the spin glass phase is
reduced and simultaneously that for ferromagnetism increases;
also, the straight lines defining the boundary of the SG phase are
the same for the three cases, since these depend only on $h_{0}$
and $\Delta$. In panel (c), ($\Delta=0.1, h_{0}=0.0, \rho=0.1$)
the upper and lower transition temperature lines ($T_{f}^{+},
T_{f}^{-}$) are shown for the PM-SG transition; the upper one
(corresponding to $T_{f}^{+}$) intersects the PM-FM line thus
defining an SG region inside the FM-phase whereas the lower line
(corresponding to $T_{f}^{-}$) does not; the same behavior appears
for other values of $h_{0}$ and $\Delta$.}
\end{center}
\end{figure}

\begin{figure}[htbp]
\begin{center}
\includegraphics*[height=0.25\textheight]{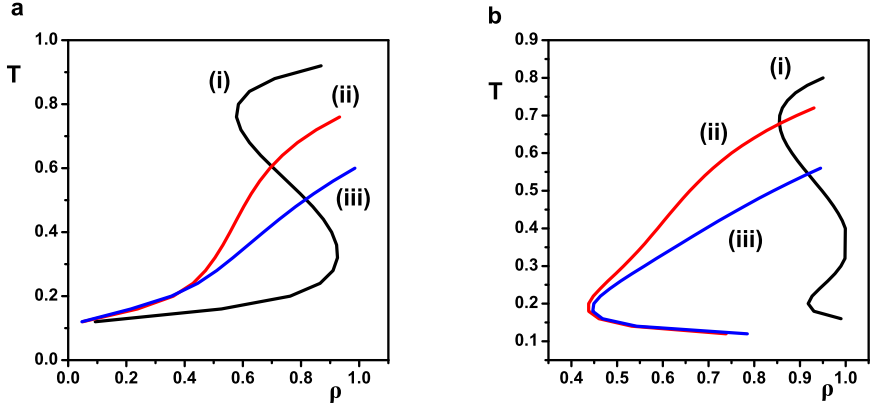}
\caption{\label{figc} (Color online) The phase diagram temperature
$T$ with respect to $\rho$ for $J_{0}=1.0$ and (a) $h_{0}=0.0$,
(b) $h_{0}=0.1$. The plots are labelled by $\Delta$, with (i)
$\Delta=0.1$, (ii) $\Delta=0.3$, (iii) $\Delta=0.5$. Because of
the limited values for $\rho$ the phase diagram is also
restricted; however, the reentrance is evident in both panels.}
\end{center}
\end{figure}

Considering also the softening to zero of the magnetization $m$,
the spin glass $q-$parameter in (\ref{mqequs}) takes the form

\begin{equation}
 q=\frac{1}{\sqrt{2\pi}} \int_{-\infty}^{\infty} e^{-z^{2}/2}
 \tanh^{2} \Biggl\{ \beta \left[ h_{0} + z (q+\Delta^{2})^{1/2} \right] \Biggr\}  \;dz   \label{q2}
\end{equation}

The ($J_{0}-T$) phase diagram appears in Fig.~\ref{figb}(a), in
the absence and presence of a random field, using equations
(\ref{mboundary1}), (\ref{mboundary2}), (\ref{templow}),
(\ref{pmsgt}) and (\ref{q2}); these ones express the
interdependence of any two quantities (mainly the temperature and
another variable) with the remaining ones considered as constants,
thus forming the respective phase diagram, in addition to the
normal one ($J_{0}-T$), Fig.~\ref{figb}(a). In this figure, the
straight lines, parallel to the $J_{0}$-axis and specifying the
boundaries of the SG phase with the aforementioned axis in the
presence of an external random magnetic field, are obtained from
Eq. (\ref{pmsgt}). If both branches of this equation are
considered, the plus-branch intersects the PM/FM boundary
enclaving a part of the FM phase inside the SG phase thus yielding
a mixed phase (see Fig.~\ref{figb}(c)), whereas the minus-branch
does not intersect the PM/FM boundary so that the mixed phase
disappears, compare Fig.~\ref{figb}(a) with Fig.~\ref{figb}(c).
Also an interesting remark is that for fixed values for $\Delta,
h_{0}$ the extent of the ferromagnetic phase increases at the
expense of the spin glass one, Fig.~\ref{figb}(b) as the
correlation $\rho$ increases; the straight line, parallel to the
$J_{0}$-axis delimiting the SG phase, is common to the three lines
because these ones correspond to the same $\Delta$ and $\rho$
values. In Fig.~\ref{figb}(c) we exhibit the phase diagram for
fixed $h_{0}$ and $\Delta$ values in the presence of both PM-SG
boundaries (upper and lower). The plus-branch is used if
$\sqrt{0.5} < T_{f} < 1$ and the minus-one if $0 < T_{f} <
\sqrt{0.5}$. The phenomenon of reentrance is also evident in
Fig.~\ref{figb} in the presence of a random magnetic field.
Moreover, one can use as independent variable any one from the
natural parameters of the system, as $\rho$; the respective "phase
diagram" appears in Fig.~\ref{figc}, in which reentrance is also
evident in both panels.

Sherrington and Kirkpatrick in their seminal paper, \cite{SK},
apart from their significant contribution to the solution of the
SG problem, their finding that the zero-temperature entropy $s(0)$
was negative had aroused a lot of debate for this peculiar
behavior until Parisi resolved this inadequacy by invoking the
replica symmetry breaking. The RS solution gives a satisfactorily
good phase diagram for the true one; however, the RS hypothesis
leads to a significant problem at low temperatures, revealed by
the negative entropy at $T=0^{\circ}K$. Using the general form for
the free energy in (\ref{rfe4}) the respective entropy as a
function of the temperature and the random field parameters
$\rho$, $h_{0}$ and $\Delta$, is

\begin{eqnarray}
s(T) = -\frac{\partial f}{\partial T} =\beta^{2} \Biggl[-\frac{(1-q)^{2}}{4} +
2q(q+\Delta ^{2}) \Biggr] - \beta m^{2} \Bigl(2J_{0}+ 3\beta \rho \Delta/4 \Bigr) +        \nonumber \\
\frac{1}{2(2\pi)^{1/2}} \int_{-\infty}^{\infty}dz \; e^{-z^{2}/2}
\ln \Bigl(4 \zeta(z)\Bigr) - \frac{h_{0} - mJ_{0}}{2T(2\pi)^{1/2}}
\int_{-\infty}^{\infty}dz \; e^{-z^{2}/2}\;\frac{\sinh(2A(z))}{\zeta(z)} -        \nonumber \\
\frac{q + \Delta^{2}}{2T^{2}(2\pi)^{1/2}}
\int_{-\infty}^{\infty}dz \; e^{-z^{2}/2} \;
\frac{2\zeta(z)\cosh(2A(z)) - \sinh^{2}(2B)}{\zeta(z)^{2}}
                                                     \;\;\;\;\;\;\;\;\; \label{entropyT}
\end{eqnarray}

where $\zeta(z) = \cosh^{2}(A(z)) - \sin^{2}B$.

The low-temperature form of the spin glass parameter $q$ is given
by $q \cong 1 - \sqrt{\frac{2}{\pi}}\frac{T}{\sqrt{1 + \Delta^{2}}}$ \cite{nishim1},
then the free energy (\ref{rfe4}) assumes the following form
in this range of temperatures

\begin{equation}
f(T,h_{0},\Delta) = -h_{0} -\sqrt{\frac{2}{\pi}}\left(1 + \frac{\Delta^{2}}{2}\right) +
\frac{\gamma T}{\pi \left(1+\Delta^{2}\right)^{1/2}}
\Bigg[1- \frac{\gamma}{2 \left(1+\Delta^{2}\right)^{1/2}}\Bigg]   \label{rfelT}
\end{equation}

where $\gamma = 1 - \frac{h_{0}^{2}}{2\left( 1+\Delta ^{2}
\right)}$; from (\ref{rfelT}) the resulting entropy per spin at
zero temperature is dependent on the random field average value
$h_{0}$ as well as on $\Delta$,

\begin{figure}[htbp]
\begin{center}
\includegraphics*[height=0.30\textheight]{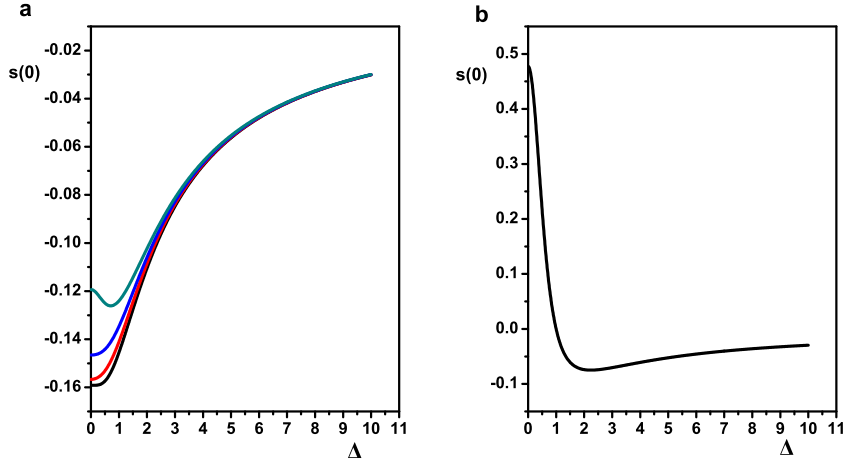}
\caption{\label{figd} (Color online) The entropy per spin at zero
temperature ($T=0^{\circ}K$) as a function of $\Delta$; panel (a)
from bottom to top $h_{0}=0.0, 0.50, 0.75, 1.00$, panel (b)
$h_{0}=2.0$.  For small $h_{0}$ values (panel (a)) it always
remains negative tending to zero as $\Delta \rightarrow \infty$,
whereas in panel (b) (strong $h_{0}$ field) it is mainly positive
tending also to zero as $\Delta \rightarrow \infty$.}
\end{center}
\end{figure}

\begin{figure}[htbp]
\begin{center}
\includegraphics*[height=0.30\textheight]{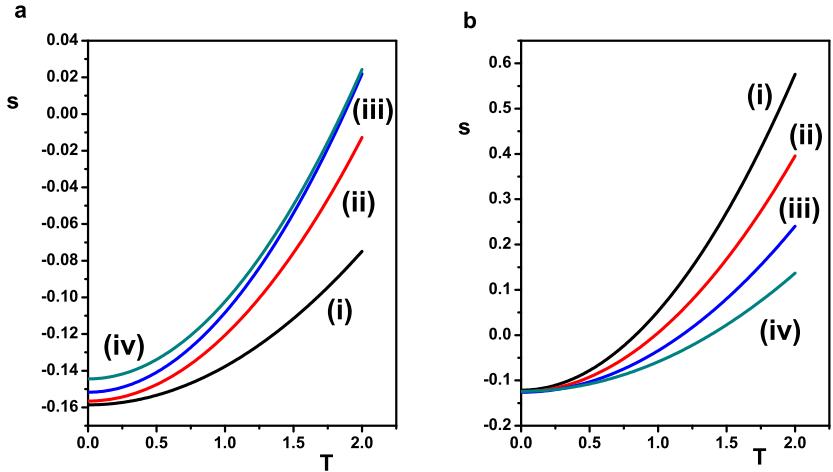}
\caption{\label{fige} (Color online) The entropy per spin as a
function of the temperature $T$ for  (a) $h_{0}=0.25$ and (b)
$h_{0}=1.00$, with $\Delta = 0.25(i), 0.50(ii), 0.75(iii),
1.00(iv)$. For low temperatures it is negative in agreement with
the Sherrington-Kirkpatrick model. In panel (a) The smaller the
$\Delta$ value the lower the plot is, whereas in panel (b) the
reverse is true.}
\end{center}
\end{figure}

\begin{figure}[htbp]
\begin{center}
\includegraphics*[height=0.25\textheight]{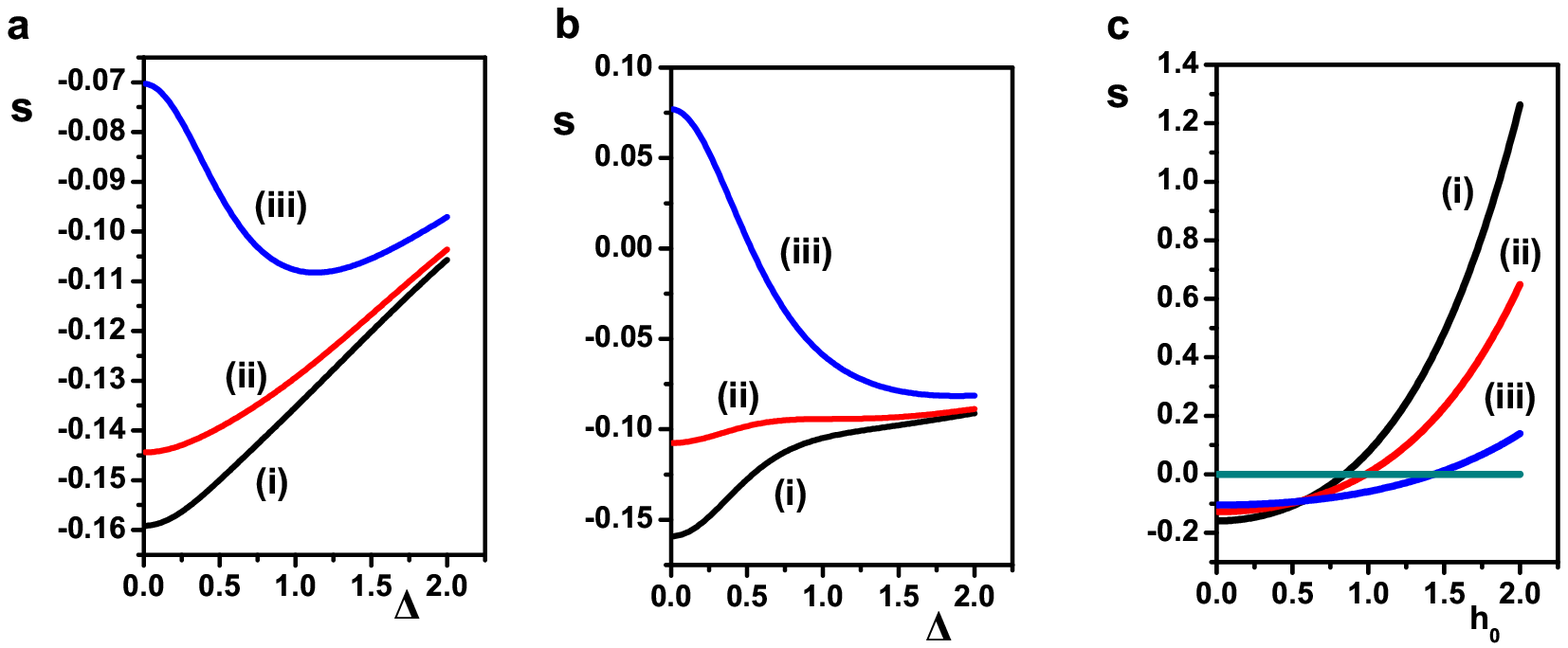}
\caption{\label{figf} (Color online) Panels (a) and (b) the
entropy per spin as a function of $\Delta$ labelled by the random
field strength $h_{0}=0.0(i), 0.5(ii), 1.0(iii)$ for temperatures,
(a) $T=0.5$, (b) $T=1.0$. Panel (c) the entropy per spin as a
function of $h_{0}$ labelled by $\Delta=0.0(i), 0.5(ii), 1.0(iii)$
for $T=1.0$.}
\end{center}
\end{figure}

\begin{equation}
s(0) =  -\frac{\gamma}{\pi \left(1+\Delta^{2}\right)^{1/2}}
\Bigg[1- \frac{\gamma}{2 \left(1+\Delta^{2}\right)^{1/2}}\Bigg]
 \label{entropyzero}
\end{equation}

remaining negative for any value of $\Delta$ and $h_{0}-$values
smaller than or equal to one, Fig.~\ref{figd} as in Ref.\cite{SK},
generalizing their result in the presence of a random magnetic
field; it tends to the zero value asymptotically as $\Delta
\rightarrow \infty$ and for $h_{0}=0, \Delta=0$,
$s(0)=-\frac{1}{2\pi}=-0.1591549$, the Sherrington-Kirkpatrick
result is recovered \cite{SK}. For $h_{0} \geq 1$ it possesses a
minimum value. However, for $h_{0}=2$, entropy $s(0)$ is mainly
positive, acquiring negative values for a small interval of
$\Delta$. From Eq. (\ref{entropyT}), entropy depends on $T,
h_{0},\Delta$; its temperature dependence appears in
Fig.~\ref{fige} representing a parabola-like curve, for specific
$h_{0}-$values and labelled by $\Delta$; for small $h_{0}$ (panel
(a)) as $\Delta$ increases the entropy increases as a function of
$T$, whereas for $h_{0}=1.0$ (panel (b)) the reverse is true. In
both cases the entropy is initially negative but, finally, becomes
positive. If, now, we consider the entropy in (\ref{entropyT}) as
a function of $\Delta$ for specific $T-$values and labelled by
$h_{0}$, then its behavior appears in Fig.~\ref{figf}($a,b$). In
this case it seems to tend to zero as $\Delta \rightarrow \infty$.
For small temperature ($T=0.5$), panel (a), and small
$h_{0}-$values it varies, nearly, linearly, but for larger
$h_{0}=1.0$ it behaves non-monotonically exhibiting a minimum
value. For higher temperature ($T=1.0$), panel (b), it behaves
non-monotonically starting from negative values for small
$h_{0}=0.0, 0.5$ whereas for higher $h_{0}=1.0$ it starts from a
positive value. In Fig.~\ref{figf}(c) the dependence of the
entropy $s$ on $h_{0}$ for $T=1.0$ and labelled by $\Delta$
appears; the entropy increases monotonically, the higher the
$\Delta-$value the less steeper is the respective entropy plot and
the three entropy branches intersect each other.

\begin{figure}[htbp]
\begin{center}
\includegraphics*[height=0.25\textheight]{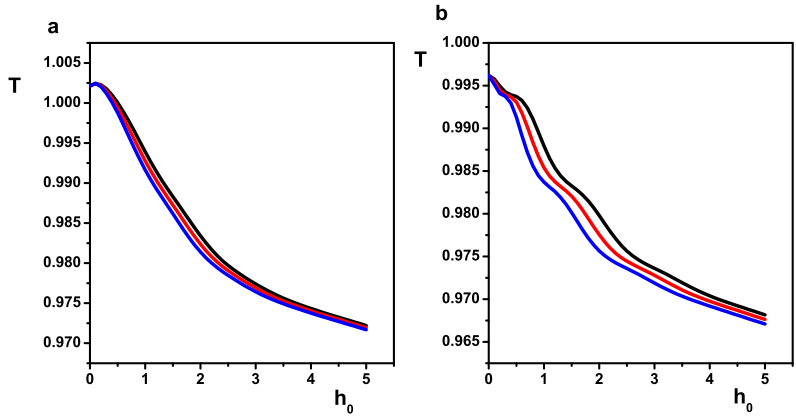}
\caption{\label{figg} (Color online) The de Almeida-Thouless line
(temperature $T$ versus the random field strength $h_{0}$),
labelled by the same values of the correlation $\rho=0.0, 0.5,
1.0$; panel (a) corresponds to $J_{0}=0.5, \Delta=0.5$; panel (b)
corresponds to $J_{0}=1.0, \Delta=1.0$; the order of the plots in
each panel is from top to bottom.}
\end{center}
\end{figure}

\begin{figure}[htbp]
\begin{center}
\includegraphics*[height=0.28\textheight]{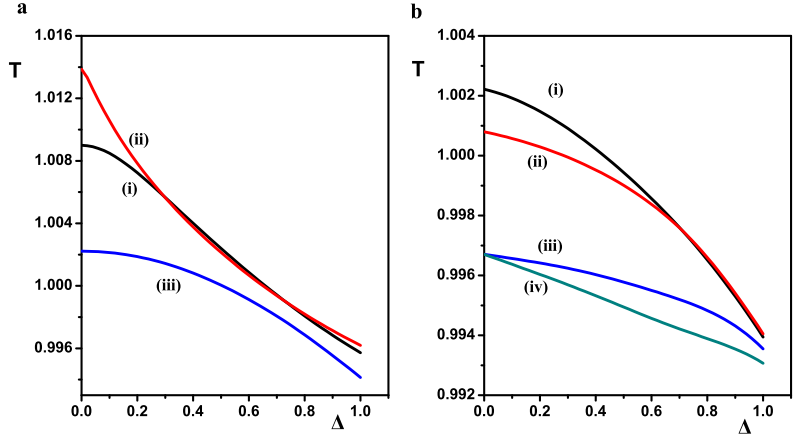}
\caption{\label{figh} (Color online) The de Almeida-Thouless line,
namely temperature $T$ versus the random field strength standard
deviation $\Delta$, labelled by the natural parameters; panel (a):
(i) $J_{0}=0.2, h_{0}=0.2, \rho=0.2$, (ii) $J_{0}=0.5, h_{0}=0.0,
\rho=0.5$, (iii) $J_{0}=0.5, h_{0}=0.5, \rho=0.0$; panel (b): (i)
$J_{0}=0.5, h_{0}=0.5, \rho=0.5$, (ii) $J_{0}=1.0, h_{0}=0.3,
\rho=0.3$, (iii) $J_{0}=1.0, h_{0}=0.5, \rho=0.2$, (iv)
$J_{0}=1.0, h_{0}=0.5, \rho=0.5$.}
\end{center}
\end{figure}

\begin{figure}[htbp]
\begin{center}
\includegraphics*[height=0.25\textheight]{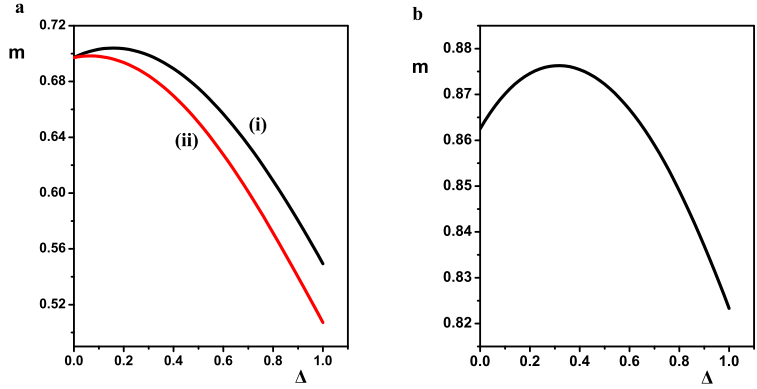}
\caption{\label{figi} (Color online) The magnetization $m$ versus
the random field strength standard deviation $\Delta$
corresponding to the temperature values of the de Almeida-Thouless
line in Fig.~\ref{figg}, labelled by the natural parameters; panel
(a): (i) $J_{0}=1.0, h_{0}=0.5, \rho=0.2$, (ii) $J_{0}=1.0,
h_{0}=0.5, \rho=0.5$,; panel (b): (i) $J_{0}=1.0, h_{0}=1.0,
\rho=1.0$.}
\end{center}
\end{figure}

\begin{figure}[htbp]
\begin{center}
\includegraphics*[height=0.19\textheight]{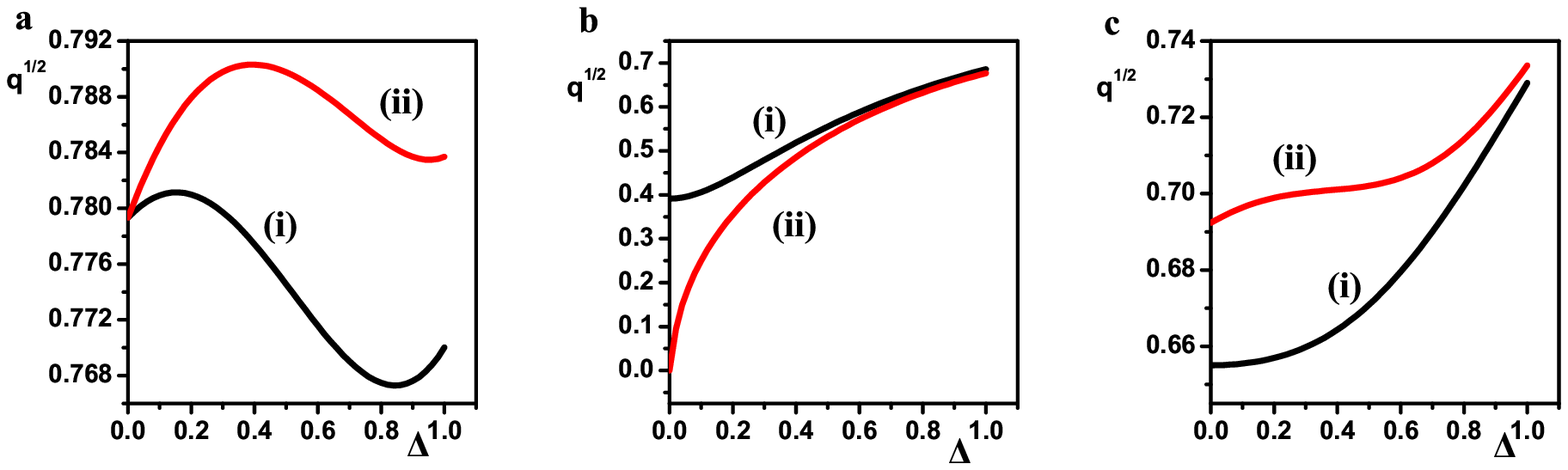}
\caption{\label{figj} (Color online) The Edwards-Anderson
parameter $q^{1/2}$ versus the random field strength standard
deviation $\Delta$ corresponding to temperature values of the de
Almeida-Thouless line in Fig.~\ref{figg}, labelled by the natural
parameters; panel (a): (i) $J_{0}=1.0, h_{0}=0.5, \rho=0.2$, (ii)
$J_{0}=1.0, h_{0}=0.5, \rho=0.5$; panel (b): (i) $J_{0}=0.2,
h_{0}=0.2, \rho=0.2$, (ii) $J_{0}=0.5, h_{0}=0.0, \rho=0.5$; panel
(c): $J_{0}=0.5, h_{0}=0.5, \rho=0.0$, (iii) $J_{0}=1.0,
h_{0}=0.3, \rho=0.3$.}
\end{center}
\end{figure}

An important issue, arisen from Sherrington-Kirkpatrick paper,
is the stability of the solution, which is reduced to the requirement the eigenvalues
of the Hessian matrix, associated with the exponential functional following the
trace in the expression (\ref{rfe2}), to be positive \cite{AT}; if at least
one of them becomes negative the symmetric solution is not correct
resulting in replica symmetry breaking. For $n\geq 1$, the eigenvalues are real,
but when analytic continuation $n \rightarrow 0$ is considered this
guarantee is lost. However, within the replica symmetric solution one
of the eigenvalues of the aforementioned Hessian matrix is negative
below a temperature defined by the functional

\begin{equation}
 T^{2} = \left(\frac{1}{2\pi}\right)^{1/2}\int_{-\infty}^{\infty}
 dz \; e^{-z^{2}/2} sech^{4}(\beta \xi(z))    \label{ATline}
\end{equation}

where $\xi(z) = h_{0} + J_{0}m + \frac{1}{2} \beta \rho \Delta m +
z\left(\Delta^{2}+q\right)^{1/2}$. Eq. (\ref{ATline}) defines a
phase boundary between the SG phase and PM (FM) phase, called de
Almeida-Thouless line (AT line), see Fig.~\ref{figg}. A
characteristic feature of Fig.~\ref{figg}(b) is that all the
AT-lines are fluctuating, whereas this phenomenon is missing from
Fig.~\ref{figg}(a). In the high temperature region the replica
symmetry is stable and the existing phase is the PM solution; at
lower temperatures the RS solution is not valid characterizing an
SG phase. The physical consequence of the AT-line is that at any
temperature $T$ below $T_{f}$ there exist a magnetic field
$H_{c}(T)$ above which the system can be described by the
mean-field equations of Sherrington-Kirkpatrick model \cite{SK},
whereas below that field a specific hypothesis concerning the
structure of the SG-phase must be done. The latter assertion
constitutes the cornerstone of the Parisi-Toulouse hypothesis
\cite{partou,vtp,monbou,fischer1,ma}. For $J_{0} \neq 0$ the AT
line, for any value of $\rho$, results as a simultaneous solution
of Eqs. (\ref{mqequs}) and (\ref{ATline}); however, for low
temperatures this line assumes a simpler form, by considering
mainly the influence of the random field as well as the
correlation $\rho$ on the behavior of the random exchange integral

\begin{equation}
 T = \frac{4}{3} \frac{1}{\sqrt{2\pi}}
\frac{\exp\left\{-\frac{1}{2} \frac{\left(h_{0} + J_{0}m +
\frac{1}{2} \rho \Delta  m \right)^{2}}{\Delta^{2}+q}
\right\} }{\left(\Delta^{2}+q\right)^{1/2}}   \label{ATline2}
\end{equation}

\begin{figure}[htbp]
\begin{center}
\includegraphics*[height=0.25\textheight]{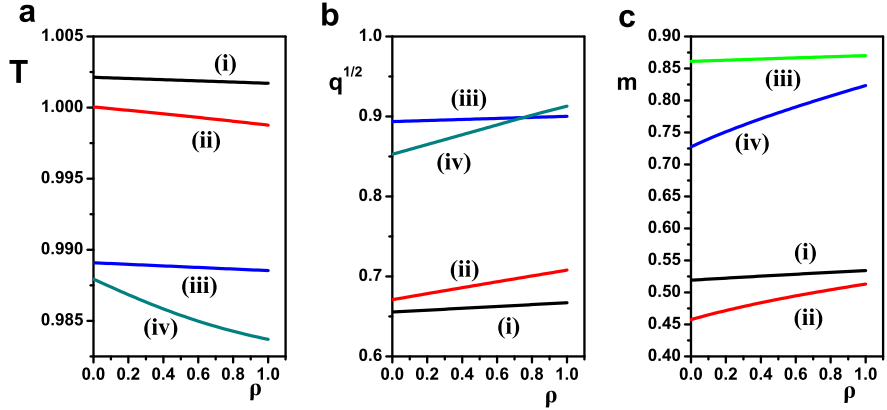}
\caption{\label{figk} (Color online) The AT profiles with respect
to the correlation $\rho$ and labeled by $J_{0}, h_{0}, \Delta$
with (i)$J_{0}=0.5, h_{0}=0.5, \Delta=0.1$, (ii)$J_{0}=0.5,
h_{0}=0.5, \Delta=0.5$, (iii)$J_{0}=1.0, h_{0}=1.0, \Delta=0.1$,
(iv)$J_{0}=1.0, h_{0}=1.0, \Delta=1.0$: (a) temperature; (b)
Edwards-Anderson parameter $q^{1/2}$; (c) magnetization .}
\end{center}
\end{figure}

However, the current system possesses more natural variables,
which influence it in one or another way, so that one can define
an "AT line" as the temperature with respect to one of these
variables besides the one corresponding to Fig.~\ref{figg}. As
such independent variable, the random field standard deviation
$\Delta$ is chosen and the respective "AT line" appears in
Fig.~\ref{figh} for specific values of the remaining natural
parameters using (\ref{ATline}), as well. Also, in Fig.~\ref{figi}
and Fig.~\ref{figj} appear the respective profiles for the
magnetization and the square root of Edwards-Anderson parameter
with respect $\Delta$ for the respective temperatures resulting
from (\ref{ATline}); these graphs display, in some cases, a non
monotonic behavior with maxima and minima. In case the correlation
$\rho$ is considered as the main control parameter, the remaining
natural parameters are fixed to specific values, then the
resulting AT profiles with respect to $\rho$ were calculated along
the AT line, Eq. (\ref{ATline}), and do not display any
significant structure as those with respect to $\Delta$, see
Fig.~\ref{figk}.

\vspace{-5mm}

\section{Conclusions and discussions}

In the current investigation, we have explored the thermodynamic
properties of the infinite-range Sherrington-Kirkpatrick Ising
spin glass model in the presence of a random field by employing a
joint Gaussian probability density function for both random
variables (the exchange integral and random field) with
correlation $\rho$ by means of the replica trick formulation.
Following the traditional route, the free energy was calculated
along the lines of the replica symmetry and, on the basis of the
saddle point method, the functional forms for the magnetization
$m$ as well as the Edwards-Anderson parameter $q$ were calculated,
which form a system of simultaneous equations. Initially, making
use of the small argument expansion, the generalized forms for the
magnetization and Edwards-Anderson parameter were found
corresponding to the respective formulae of the PaT hypothesis. We
have also found a generalized form for the SK zero-temperature
value for $J(T=0)=\sqrt{\frac{\pi}{2}}$ by proposing one depending
on $h_{0}$ and $\Delta$; this expression is non-monotonic as a
function of $\Delta$ and labelled by $h_{0}$. The temperature,
delimiting the SG phase, has been calculated analytically
depending only on $\Delta$ and $h_{0}$; from its functional
expression it can be determined the possibility of the existence
or not of the SG phase. The structure of the phase diagram ($T$
with respect to $J_{0}$) is more rich and as the correlation
increases the extent of the SG phase is reduced, thus increasing
that of the FM phase; in addition to the former phase diagram, the
random system possesses more ones, i.e., the temperature $T$ with
respect to $\Delta$ and $h_{0}$; however, the random system
possesses a plethora of natural variables ($T, J_{0}, h_{0},
\Delta, \rho$), which affect the behavior of the system, implying
that one can plot the "phase diagram" by using a pair of these
variables keeping the remaining ones fixed.

The zero temperature entropy ($s(0) = -\frac{1}{2\pi}$) was
generalized with the new expression depending on $h_{0}$ and
$\Delta$; $s(0)$ as a function of $\Delta$ initially is monotonic
but it becomes non-monotonic as well as positive, also the entropy
as a function of temperature is represented by a  parabola-like
curve. A similar complex variation presents the de
Almeida-Thouless line ($T$ vs $h_{0}$) so that as the correlation
$\rho$ increases the SG region is reduced whereas that of the FM
phase increases. We have also calculated the magnetization as well
as the square root of the EA parameter profiles with respect to
$\Delta$ for the respective de Almeida-Thouless line temperature;
in both cases the respective profiles are mainly non-monotonic.

Although the formulation is along the lines of the mean field
theory, this description cannot be considered as a trivial one, in
that it is very complex in its description requiring complicated
mathematics for the description of the low temperature phase - the
spin glass phase. The SG phenomena are essentially dynamic
critical ones, and more developments are needed to give us a
better understanding of the spin glasses. In this case we shall
rely on short range models concentrating on Monte Carlo
simulations and the Renormalization Group methods. One may assert
that the present situation about spin glasses is successful in the
sense that a qualitative understanding of the SG phase has been
achieved by introducing innovative methods in Statistical Physics
as the Replica Symmetry Breaking as well as the Ultrametricity,
since the Replica Theory provides a powerful approach to the
complex low temperature phase of these systems, which is described
in terms of a "Replica Symmetry Broken" solution. However, an
important open issue is to understand whether the RSB scenario
also holds beyond the mean-field approximation. Still, the glassy
behaviour is observed at low temperatures and the phenomenology
much resembles the one of some mean-field spin glass models. For
this reason, concepts and techniques from spin glasses have been
widely applied to investigate glassy behaviour in these systems.

The current investigation shall be extended towards the Replica
Symmetry Breaking formulation.

\noindent
\ack{This research was supported by the Special Account
 for Research Grants of the University of Athens $\left(E\Lambda
 KE\right)$ under Grant No. 70/4/4096.}

\end{document}